\documentclass[runningheads]{llncs}
\usepackage{graphicx}
\usepackage{hyperref}

\usepackage{times}
\usepackage{latexsym}
\usepackage{amsmath}
\usepackage{amssymb}
\usepackage{multirow}
\usepackage{url}
\usepackage{booktabs}
\usepackage{pgfplots}
\usepackage{graphicx}
\usepackage{subfig}
\usepackage{wrapfig}
\usepackage{caption}
\usepackage{color}
\usepackage{floatrow}
\usepackage{microtype}
\usepackage{tikz}
\usepackage{listing}
\usepackage{siunitx}
\usepackage{placeins}

\usepackage{lscape}
\usepackage{cleveref}
\usepackage{array}
\usepackage{tabulary}
\newcolumntype{C}[1]{>{\centering\arraybackslash}p{#1}}
\newcolumntype{L}[1]{>{\raggedright\arraybackslash}p{#1}}
\newcolumntype{R}[1]{>{\raggedleft\arraybackslash}p{#1}}

\newcommand{\domain}{\ensuremath{D}}
\newcommand{\knn}{\mbox{$k$-NN} }
\newcommand{\topk}{top-\textit{k}}
\newcommand{\ttt}{\texttt}
\newcommand{\tfidf}{TF$\times$IDF }

\newcommand{\knnns}{\mbox{$k$-{NN}}}

\newcommand{\yahoons}{Yahoo}
\newcommand{\yahoo}{\yahoons{} }

\newcommand{\second}[1]{\SI{#1}{\second}}

\newif\ifProduction

\ifProduction
\newcommand{\leoinline}[1]{}

\else
\newcommand{\leoinline}[1]{\textbf{\color{red}\small #1}}

\fi

\begin{document}
\title{Accurate and Fast Retrieval for Complex Non-Metric Data via Neighborhood Graphs\thanks{
This work was accomplished while Leonid Boytsov was a PhD student at CMU.
Authors gratefully acknowledge the support by the NSF grant \#1618159: ``Matching and Ranking via Proximity Graphs: Applications to Question Answering and Beyond''.}
}
\titlerunning{Searching Non-Metric Data via Neighborhood Graphs}
%
%
\author{Leonid Boytsov\inst{1}
\and
Eric Nyberg\inst{2}
}
\authorrunning{L. Boytsov and E. Nyberg}
%
\institute{Carnegie Mellon University, Pittsburgh, PA, {srchvrs@cs.cmu.edu}
\and
Carnegie Mellon University, Pittsburgh, PA, {enh@cs.cmu.edu}
}
\maketitle              
\begin{abstract}
We demonstrate that a graph-based search algorithm---relying 
on the construction of an approximate neighborhood graph---can directly work with
challenging \emph{non-metric} and/or \emph{non-symmetric} distances without resorting to metric-space mapping and/or distance symmetrization, which, in turn, lead to substantial performance degradation.
Although the straightforward metrization and symmetrization is usually ineffective,
we find that constructing an index using a modified, e.g., symmetrized, distance can improve performance. 
This observation paves a way to a new line of research
of designing index-specific graph-construction distance functions.
This is an archival version, the publisher's version is available
at \href{https://link.springer.com/chapter/10.1007/978-3-030-32047-8_12}{Springer.com}.
\end{abstract}
\section{Introduction and Problem Definition}
In this paper we focus on $k$ nearest neighbor (\knnns) search,
which is a widely used computer technology with
applications in machine learning, data mining, information retrieval,
and natural language processing.
Formally, we assume to have a possibly \emph{infinite} domain containing
objects $x$, $y$, $z$, \ldots, which are commonly called data points or simply points.
The domain---sometimes called a \emph{space}---is equipped with 
with a \emph{distance function} $d(x,y)$,
which is used to measure dissimilarity of objects $x$ and $y$.
The value of $d(x,y)$ is interpreted as a degree of dissimilarity.
The larger is $d(x,y)$, the more dissimilar points $x$ and $y$ are.

Some distances are non-negative and become zero
only when $x$ and $y$ have the highest possible degree of similarity.
The \emph{metric} distances are additionally symmetric and satisfy the triangle inequality.
However, in general, we do not impose any restrictions on the value of the distance function (except that smaller values represent more similar objects).
Specifically, the value of the distance function can be negative and
negative distance values indicate higher similarity than positive ones.

We further assume that there is a data \emph{sub}set \domain\ containing a \emph{finite}
number of domain points and a set of queries that belongs to the domain but not to \domain.
We then consider a standard \topk\  retrieval problem. 
Given a query $q$
it consists in finding $k$ data set points $\{x_i\}$ with smallest values of distances
to the query among all data set points (ties are broken arbitrarily). 
Data points $\{x_i\}$ are called \emph{nearest neighbors}.
A search should return $\{x_i\}$ in the order of increasing distance to the query.
If the distance is not symmetric, two types of queries can be considered:
\emph{left} and \emph{right} queries.
In a \emph{left} query, a data point compared to the query is always
the first (i.e., the left) argument of $d(x,y)$.
Henceforth, for simplicity of exposition we consider only the case of left queries.

Exact methods degenerate to a brute-force search for just a dozen of dimensions 
\cite{weber1998quantitative}.
Due to diversity of properties, non-metric spaces lack \emph{common} and \emph{easily identifiable} structural properties such as the triangle inequality.
There is, therefore, little hope that \emph{fully} generic exact search methods can be devised.
Thus, we focus on the \emph{approximate} version of the problem where 
the search may miss some of the neighbors, but it may not change the order.
The accuracy of retrieval is measured via recall (equal to the 
average fraction of neighbors found).
We cannot realistically devise fast exact methods, 
but we still hope that our approximate methods are quite \emph{accurate}
having a recall close to 100\%.

There has been a staggering amount of effort invested in designing new and improving existing \knn search algorithms (see
e.g., \cite{chavez2001searching,Samet2005,Skopal2011,wang2014hashing}).
This effort has been placed disproportionately on techniques for \emph{symmetric}
\emph{metric} distances, in particular, on search methods for the Euclidean space.
Yet, search methods for challenging non-symmetric and non-metric spaces received very \emph{little} attention.
A \emph{filter-and-refine} approach is a common way to deal with an unconventional distance.
To this end one would map data to a low-dimensional Euclidean space.
The goal is to find a mapping without large distortion of the original similarity measure \cite{Jacobs_et_al:2000,Hjaltason_and_Samet:2003}.
Jacobs et al.\ \cite{Jacobs_et_al:2000} review various projection methods
and argue that such a coercion is often against the nature of a similarity measure,
which can be, e.g., intrinsically non-symmetric.
Yet, they do not provide experimental evidence.
We fill this gap and demonstrate that both metric learning and distance symmetrization are, indeed, suboptimal approaches.

Alternatively the metric distance can be learned from scratch \cite{DBLP:journals/corr/BelletHS13}.
In that, Chechik et al.~\cite{DBLP:journals/jmlr/ChechikSSB10} contended that in the task
of distance learning enforcing symmetry and metricity is useful only as a means to prevent overfitting to a small training set.
However, when training data is abundant, it can be more efficient and more accurate to learn
the distance function in an unconstrained bilinear form.
Yet, this approach does not necessarily results in a symmetric metric distance \cite{DBLP:journals/jmlr/ChechikSSB10}.
We, in turn, demonstrate that a graph-based retrieval algorithm---relying 
on the construction of approximate neighborhood/proximity graphs---can deal with
challenging non-metric distances directly without resorting to a low-dimensional mapping or full symmetrization.
In that, unlike prior work \cite{Naidan2015,ponomarenko2014comparative},
as we show in \S~\ref{SectionExperFilterBF}, several of our distances
are substantially non-symmetric.

Whereas the filter-and-refine symmetrization approach is detrimental, 
we find that constructing an index using the symmetrized distance can improve results. 
Furthermore, we show that the index construction algorithm can be quite sensitive to 
the order of distance function arguments.
In most cases, changing the argument order is detrimental. However,
this is not a universal truth:
Quite surprisingly, we observe small improvements in some cases
by building the graph using the argument-reversed distance function.
We believe this observations motivates the line of research
 to design indexing distance functions---different from
original distance functions---that result in better performance. 
The remaining paper contains the description of employed 
retrieval algorithms and related experimental results.

\section{Methods and Materials}
\subsection{Retrieval Algorithms}\label{SectionAlgo}

We consider two types of retrieval approaches: the filter-and-refine method using brute-force search and
indexing using the graph-based retrieval method Small World Graph (SW-graph) \cite{malkov2014approximate}.
In the filter-and-refine approach, 
we use a proxy distance to generate a list 
of $k_c$ candidate entries (closest to the query with respect to the proxy distance) via
the brute-force, i.e., exhaustive, search. 
For $k_c$ candidate entries $x_i$ we compute the true distance values
$d(x_i, q)$---or $d(q,x_i)$ for right queries---and select $k$ closest entries.

The filter-and-refine approach can be slow even if the proxy distance is quite cheap~\cite{Naidan2015},
whereas indexing can dramatically speed up retrieval.
In particular, state-of-the-art performance can be achieved by using graph-based
retrieval methods, which rely on the construction of an exact or approximate \emph{neighborhood} graph
(see, e.g., \cite{aumuller2019ann,Naidan2015}).
The neighborhood graph  is a data structure in which data points are associated
with graph nodes and sufficiently close nodes are connected by edges.
A search algorithm 
is a graph-traversal routine 
exploiting  a property ``the closest neighbor of my closest neighbor is my neighbor as well.'' 
The neighborhood graph is often defined as a directed graph \cite{hajebi2011fast,dong2011efficient},
where the edges go from a vertex to its neighbors (or vice versa),
but undirected edges have been used too  \cite{malkov2014approximate,DBLP:journals/corr/LiZSWZL16}
(undirected nodes were also quietly introduced in kgraph\footnote{\url{https://github.com/aaalgo/kgraph}}).
In a recent study, 
the use of undirected neighborhood graphs
lead to a better performance \cite{DBLP:journals/corr/LiZSWZL16}.

Constructing an \emph{exact} neighborhood graph is hardly feasible for a large high-dimensional data set,
because, in the worst case, the number of distance computations is $O(n^2)$,
where $n$ in the number of data points. 
An \emph{approximate} neighborhood graph can be constructed substantially more efficiently
\cite{malkov2014approximate,dong2011efficient}.
To improve performance, one can use various graph pruning methods \cite{DBLP:journals/corr/LiZSWZL16,2016arXiv160309320M,DBLP:conf/cvpr/HarwoodD16}:
In particular, it is not useful to keep neighbors that are close to each other \cite{DBLP:journals/corr/LiZSWZL16,DBLP:conf/cvpr/HarwoodD16}.

Neighborhood graphs have a long history.
Toussaint published a pioneering paper where he introduced 
neighborhood graphs on the plane in 1980 \cite{toussaint1980relative}.
Arya and Mount were first to apply neighborhood graphs
to the problem of \knn\  search in a high-dimensional space \cite{arya1993approximate}.
Houle and Sakuma proposed the first hierarchical, i.e., multi-layer,
variant of the neighborhood graph called SASH, where data points at layer $i$
are connected only to the nodes at layer $i+1$ \cite{DBLP:conf/icde/HouleS05}.
Malkov and Yashunin proposed  an efficient multi-layer neighborhood-graph
method called a Hierarchical Navigable Small World (HNSW) \cite{2016arXiv160309320M}.
It is a generalization and improvement of the previously proposed method navigable Small World (SW-graph) \cite{malkov2014approximate},
which has been shown to be quite efficient in the past  \cite{malkov2014approximate,Naidan2015}

Although there are different approaches to construct a neighborhood graphs,
all retrieval strategies known to us rely on a simple semi-greedy graph-traversal
algorithm with (possibly) multiple restarts.
Such an algorithm keeps a priority queue of elements, 
which ranks candidates in the order of increasing distance to the query. 
At each step, the search retrieves one or more elements from the queue
that are closest to the query  and explores their neighborhoods.
Previously unseen elements may be added to the queue.
For a recent experimental comparison of several retrieval approaches see \cite{DBLP:journals/corr/TellezRCG17}.

Although, HNSW is possibly the best retrieval method for generic distances \cite{2016arXiv160309320M,DBLP:journals/corr/LiZSWZL16},
in our work we use a modified variant of SW-graph, where retrieval starts from a single point (which is considerably more
efficient compared to multiple starting points).
The main advantage of HNSW over the older version of SW-graph is due to 
(1) introduction of pruning heuristics,
(2) using a single starting point during retrieval.
We want to emphasize that comparison of HNSW against SW-graph in \cite{2016arXiv160309320M} is not completely fair,
because it basically uses an undertuned SW-graph.
Furthermore, gains from using a hierarchy of layers are quite small: see Fig. 3-5 from \cite{2016arXiv160309320M}.
At the same time pruning heuristics introduce another confounding factor in measuring the effect of distance symmetrization (and proxying),
because symmetrization method used in the pruning approach can be different from the symmetrization method used by \knn\ search
employed at index time.
Thus---as we care primarily about demonstrating usefulness (or lack thereof) of different distance modifications during construction of the graph
rather than merely achieving maximum retrieval efficiency---we experiment with a simpler retrieval algorithm SW-graph.
The employed algorithm has three main parameters.
Parameter \ttt{NN} influences (but does not define directly) the number of neighbors in the graph.
Parameters \ttt{efConstruction} and \ttt{efSearch} define the depth of the priority queue
used during index and retrieval stages, respectively.

\begin{table}[tb]
\centering
\caption{Data sets\label{TableData}}
\begin{tabular}{l@{\hspace{1em}}l@{\hspace{1em}}l@{\hspace{1em}}L{6cm}}
\toprule
Name & max. \# of rec. & Dimensionality   & Source   \\ \midrule

RandHist-$d$  & $0.5\times10^6$  & $d \in \{8,32\}$ & Histograms sampled uniformly from a simplex   \\
RCV-$d$ & $0.5\times10^6$ & $d \in \{8,128\}$  & $d$-topic LDA \cite{blei2003latent} RCV1 \cite{DBLP:journals/jmlr/LewisYRL04} histograms \\
Wiki-$d$ & $2\times10^6$  & $d \in \{8,128\}$  & $d$-topic LDA \cite{blei2003latent} Wikipedia histograms  \\
Manner & $1.46\times10^5$ & $1.23\times 10^5$  &  Question and answers from L5 collection in Yahoo WebScope
\\ \bottomrule
\end{tabular}
\end{table}

\begin{table}[tb]
\centering
\caption{Distance Functions}
\label{TableDist}
\begin{tabular}{@{}L{3cm}c@{\hspace{0.5em}}L{4cm}@{}}
\toprule
Denotation/Name & d(x,y) & Notes \\ \midrule

Kullback-Leibler diverg. (KL-div.) \cite{kullback1951} & $\sum\limits_{i=1}^m  x_i\log{\dfrac{x_i}{y_i}}$ &
\\

Itakura-Saito distance \cite{itakura1968analysis} & $\sum\limits_{i=1}^m \left[ \frac{ x_i}{y_i} - \log \frac{x_i}{y_i}  -1 \right]$ &
\\

R\'{e}nyi diverg. \cite{rrnyi1961measures} &  $\frac{1}{\alpha-1}\log\left[\sum\limits_{i=1}^m x_i^\alpha y_i^{1-\alpha}\right]$,
$0 < \alpha < \infty$
&
We use $\alpha\in{0.25,0.75,2}$
\\

BM25 similarity \cite{Robertson2004} & 
$-\sum_{x_i=y_i} \textrm{TF}_q(x_i) \cdot \textrm{TF}_d(y_i) \cdot \textrm{IDF}(y_i)$
& $\textrm{TF}_q(x)$ and $\textrm{TF}_d(y)$ are (possibly scaled)
term frequencies in a query and document.
\\

\bottomrule
\end{tabular}
\end{table}

\subsection{Data sets and Distances}\label{SectionData}

In our experiments, we use the following distances (see Table \ref{TableDist}):
KL-divergence,  the Itakura-Saito distance,
the R\'{e}nyi divergence, and BM25 similarity \cite{Robertson2004}.
The first three distances are statistical distances defined
over probability distributions.
Statistical distances in general and, KL divergence in particular, play an important role in ML \cite{Cayton2008,sutherland2016scalable}. 
Both the KL-divergence and the Itakura-Saito distances
were used in prior work \cite{Cayton2008}.
BM25 similarity is a popular and effective similarity metric
commonly used in information retrieval. 
It is a variant of a \tfidf similarity computed as 
\begin{equation}\label{EqBM25}
\sum_{x_i=y_i} \textrm{TF}_q(x_i) \cdot \textrm{TF}_d(y_i) \cdot \textrm{IDF}(y_i),
\end{equation}
where $\textrm{TF}_q(x)$ and $\textrm{TF}_d(y)$ are 
term frequencies of terms $x$ and $y$ in a query and a document, respectively.
$\textrm{IDF}$ is an inverse document frequency (see \cite{Robertson2004} for more details).
When we use BM25 as a distance, we take the negative value of this similarity function.
Although BM25 is expressed as an inner product between
query and document \tfidf vectors, 
this distance is \emph{not} symmetric.
Term frequencies are computed differently for queries and documents and the value
of the similarity normally changes when we swap function arguments.

The R\'{e}nyi divergence is a single-parameter family of distances,
which are not symmetric when the parameter $\alpha \ne 0.5$.
By changing the parameter we can vary the degree of symmetry.
In particular, large values of $\alpha$ as well as close-to-zero values
result in highly non-symmetric distances.
This flexibility allows us to stress-test retrieval methods
by applying them to challenging non-symmetric distances.

The data sets are listed in Table~\ref{TableData}.
Wiki-$d$ and RCV-$d$
data sets consists of dense vectors of topic histograms with $d$ topics.
RCV-$d$ set are created by Cayton \cite{Cayton2008} 
from the RCV1 newswire collection \cite{DBLP:journals/jmlr/LewisYRL04}
using the latent Dirichlet allocation (LDA) method \cite{blei2003latent}. 
These data sets have only 500K entries.
Thus, we created 
larger sets from Wikipedia following a similar methodology.
RandHist-$d$ is a synthetic set of topics sampled uniformly 
from a $d$-dimensional simplex.

The Manner data set is a collection of \tfidf vectors
generated from data set L5 in \yahoo WebScope\footnote{\url{https://webscope.sandbox.yahoo.com}}.
L5 is a set of manner, e.g., how-to, questions posted on the \yahoo answers webite
together with respective answers.
Note that we keep only a single best answer---as selected by a community member---for each question.

\section{Experiments}

We carry out two experimental series.
In the first series, we test the efficacy of the filter-and-refine
approach (using collection subsets) where the distance function is obtained via
metrization or symmetrization of the original distance.
One of the important objectives of this experimental series
is to demonstrate that unlike some prior work \cite{Naidan2015,ponomarenko2014comparative}
we  deal with \emph{substantially} \emph{non}-symmetric data.
In the second series, 
we carry out a fully-fledged retrieval experiment using SW-graph \cite{malkov2014approximate} with different index- and query-time
symmetrization approaches.
Overall, we have 31 combination of data sets and distance functions (see \S~\ref{SectionData}).
However, due to space limitations, we had to omit some experimental results and minor setup details. A fuller description is available in \S 2.3.2 of the unpublished tech report \cite{boytsov2018efficient}.

\subsubsection{Proxying Distance via Metrization and Symmetrization}
\label{SectionExperFilterBF}
In this section, we use a proxy distance function to generate a list of $k_c$
candidates, which are compared directly to the query. 
The candidate generation step employs an \emph{exact} brute-force \knn search with the proxy distance.
On one hand, the larger is $k_c$, the more likely we find all true nearest neighbors.
On the other hand, increasing $k_c$ entails a higher computational
cost. We consider two types of proxy distances: a learned distance (which is a metric
in four out of five cases),
and a symmetrized version of the original non-symmetric distance. 

\paragraph{Distance learning} 
We considered five approaches to learn
a distance and a pseudo-learning approach where we simply use the Euclidean
$L_2$ distance as a proxy. 
Computing $L_2$ between data points is a strong baseline,
which sometimes outperforms true distance learning methods,
especially for high-dimensional data.
Four of the distance-learning methods  
\cite{DBLP:conf/nips/WeinbergerBS05,DBLP:conf/icml/DavisKJSD07,DBLP:conf/icml/QiTZCZ09,DBLP:conf/icdm/LiuGZJW12} learn a global linear transformation of the data,
which is commonly referred to as  the Mahalanobis metric learning. 
The value of the $L_2$ distance between transformed vectors is used as a proxy distance function.
The learned distance, is clearly a metric.
We also use a non-linear Random Forest Distance (RFD) method that employs
a random-forest classifier \cite{DBLP:conf/kdd/XiongJXC12} and produces
generally non-metric, but symmetric, distance.
Note that we do not learn a distance function for the Manner data set that 
contains extremely high dimensional sparse \tfidf vectors.

In all cases, the distance is trained as a classifier
that learns to distinguish between close and distant data points.
More specifically, we create sets of positive and negative examples.
A positive example set contains pairs of points that should be treated as similar, i.e., near points,
while the negative example set contains pairs of points that should be treated as dissimilar ones.
The underlying idea is to learn a distance that (1) pulls together points from the positive example
set and (2) pushes points from the negative example set apart. More details are given in \cite{boytsov2018efficient}.

\paragraph{Symmetrization}\label{ParaSymmDesc}
Given a non-symmetric distance, there are two folklore approaches to make it symmetric,
which use the value of the original distance $d(x,y)$ as well as the value of the
distance function obtained by reversing arguments: $ d_{\textrm{reverse}}(x,y)=d(y,x) $.
Informally, we call the latter an \emph{argument-reversed} distance.
In the case of an average-based symmetrization, we compute the symmetrized distance
as an average of the original and argument-reversed distances:
\begin{equation}\label{EqGenSymmetrAvg}
d_{\textrm{sym}}
=\frac{d(x,y) + d_{\textrm{reverse}}(x,y)}{2}
=\frac{d(x,y) + d(y,x)}{2}
\end{equation}
In the case of a min-based symmetrization, we use their minimum:
\begin{equation}\label{EqGenSymmetrMin}
d_{\textrm{sym}}=\min\left(d(x,y) , d_{\textrm{reverse}}(x,y)\right) =
\min\left(d(x,y) , d(y,x)\right)
\end{equation}

Symmetrization techniques given by Eq.~(\ref{EqGenSymmetrAvg}) and Eq.~(\ref{EqGenSymmetrMin})
are suboptimal in the sense that a \emph{single} computation of the symmetrized distance
entails \emph{two} computations of the original distance.
We can be more efficient when a distance function permits a more \emph{natural} symmetrization, in particular, 
in the case of BM25 (see Eq.~\ref{EqBM25})
we can compute the query term frequency using the same formula as the 
document term frequency. 
Furthermore, we can ``share'' a value of $\textrm{IDF}_i$ between
the query and the document vectors by ``assigning'' each vector the value $\sqrt{\textrm{IDF}_i}$.
Although the resulting function is symmetric, it is not equivalent to the original BM25.
More formally, in this ``shared'' setting a query vector is represented by the values 
$\textrm{TF}(x_i)\cdot\sqrt{\textrm{IDF}(x_i)}$,
whereas a document vector
is represented by the values $\textrm{TF}(y_i) \cdot  \sqrt{\textrm{IDF}(y_i)}$.
The pseudo-BM25 similarity is computed as the inner product between query and document vectors 
in the following way:
\begin{equation}\label{EqSymmNatural}
 d(x,y)=-\sum_{x_i=y_i} \left( \textrm{TF}(x_i) \sqrt{\textrm{IDF}(x_i)} \right) \cdot \left(\textrm{TF}(y_i) \sqrt{\textrm{IDF}(y_i)} \right)
\end{equation}

{
\begin{table}[tb]
\caption{Loss of effectiveness due to symmetrization and distance learning
for 10-NN search (using at most 200K points for distance learning and at most 500K points for symmetrization)\label{TableSymmMetrLearn}}
\small
\begin{tabular}{@{}l@{\hspace{1em}}l@{\hspace{1em}}l@{\hspace{1em}}l@{\hspace{1em}}l@{\hspace{1em}}l@{}}
\toprule
Data set & Distance & \multicolumn{2}{l}{Symmetrization} & \multicolumn{2}{c}{\begin{tabular}[c]{@{}c@{}}Distance \\ learning\end{tabular}} \\ \midrule
 &  & \multicolumn{1}{c}{\begin{tabular}[c]{@{}c@{}}$k_c$ \\ (cand. $k$)\end{tabular}} & \begin{tabular}[c]{l@{}}Recall \\ reached\end{tabular} & \multicolumn{1}{c}{\begin{tabular}[c]{@{}c@{}}$k_c$ \\ (cand. $k$)\end{tabular}} & \multicolumn{1}{c}{\begin{tabular}[c]{@{}c@{}}Recall \\ reached\end{tabular}} \\ \midrule
Wiki-8 & Itakura-Saito & 20 & 99 & 2560 & 99 \\
Wiki-8 & KL-div. & 40 & 99 & 640 & 99 \\
Wiki-8 & R\'{e}nyi div. $\alpha=0.25$ & 20 & 100 & 640 & 100 \\
Wiki-8 & R\'{e}nyi div. $\alpha=2$ & 20 & 99 & 640 & 99 \\ \midrule
RCV-128 & Itakura-Saito & 80 & 99 & 20480 & 58 \\
RCV-128 & KL-div. & 40 & 100 & 20480 & 94 \\
RCV-128 & R\'{e}nyi div. $\alpha=0.25$ & 80 & 100 & 5120 & 99 \\
RCV-128 & R\'{e}nyi div. $\alpha=2$ & 80 & 99 & 20480 & 66 \\ \midrule
Wiki-128 & Itakura-Saito & 20 & 99 & 20480 & 80 \\
Wiki-128 & KL-div. & 40 & 99 & 20480 & 99 \\
Wiki-128 & R\'{e}nyi div. $\alpha=0.25$ & 160 & 99 & 5120 & 99 \\
Wiki-128 & R\'{e}nyi div. $\alpha=2$ & 80 & 99 & 20480 & 87 \\ \midrule
RandHist-32 & Itakura-Saito & 5120 & 96 & 20480 & 99 \\
RandHist-32 & KL-div. & 160 & 100 & 2560 & 99 \\
RandHist-32 & R\'{e}nyi div. $\alpha=0.25$ & 20 & 100 & 1280 & 100 \\
RandHist-32 & R\'{e}nyi div. $\alpha=2$ & 2560 & 99 & 20480 & 100 \\ \midrule
Manner & BM25 & 1280 & 100 & N/A & N/A \\ \bottomrule
\end{tabular}
\end{table}
}

\paragraph{Discussion of Results}

All the code in this section is implemented in Python.
Thus, for efficiency reason, we limit the number of data points
to 200K in the symmetrization experiment and to 500K in the
distance learning experiment.
Experimental results for $k=10$ are presented in Table~\ref{TableSymmMetrLearn},
where we measure how many candidates $k_c$ we need to achieve a nearly perfect recall with respect to the original distance (we
test all $k_c=k \cdot 2^i$, $i\le 7$).
We employ several symmetrization and distance learning methods:
Yet, in the table, we show only the best recall for a given $k_c$.
More specifically, we post the first $k_c$ for which recall
reaches 99\%. If we cannot reach 99\%, we post the maximum recall
reached. We omit most low-dimensional results, because they are similar
to Wiki-8 results (again, see \cite{boytsov2018efficient} for a more detailed report).

From Table~\ref{TableSymmMetrLearn} 
we can immediately see that distance learning results in a much
worse approximation of the original distance than symmetrization.
For high-dimensional data, it is not always possible to achieve 
the recall of 99\% for 10-NN search. 
When it is possible we need to retrieve from one thousand to 20 thousand candidate entries! Even for the low-dimensional Wiki-8 data set,
achieving such high recall requires at least 640 candidate entries.
We conclude that using distance learning is not a promising direction,
because retrieving that many candidate entries accurately is hardly possible without resorting to the brute 
force search with the proxy distance (which is, in turn, not efficient). 

In contrast, in the case of symmetrization, the number of required
candidate entries is reasonably small except for Manner and RandHist-32
data sets. We, therefore, explore various symmetrization approaches
in more details in the following section. 
Also note that KL-divergence can be symmetrized with little loss in accuracy,
i.e., on the histogram-like data KL-divergence is only mildly non-symmetric.
There is prior work on non-metric \knn\  search that demonstrated 
good results specifically for KL-divergence \cite{Naidan2015,ponomarenko2014comparative}
for Wiki-$d$ and RCV-$d$ data sets.
However, as our experiments clearly show, this work does not use
a substantially non-symmetric distance.

\subsubsection{Experiments with Index- and Query-Time Symmetrization for SW-graph}\label{SectionExperSymmetrSW}
In this section, we evaluate the effect of the distance symmetrization in two scenarios (for 10-NN search):
\begin{itemize}
\item 
A symmetrized distance is used for \emph{both} indexing and retrieval.
We call this a  full symmetrization scenario.
The search procedure is carried out using an SW-graph index \cite{malkov2014approximate} (see \S~\ref{SectionAlgo}).
This search generates a list of $k_c$ candidates.
Then, candidates are compared exhaustively with the query.
This filter-and-refine experiment is analogous to the previous-subsection experiments
except here we use approximate instead of the exact brute-force search.
\item
The second scenario relies on a partial, i.e., index-time only, symmetrization.
Specifically, the symmetrized distance is used only to construct a proximity/neighborhood graph via SW-graph.
Then, the search procedure uses the \emph{original}, \emph{non-symmetrized} distance to ``guide''
the search through the proximity graph.
\end{itemize}

Overall, we have 31 combinations of data sets and distances, 
but in this paper we present the results for most interesting cases (again see \cite{boytsov2018efficient} for a complete set of plots).
We randomly split data three times into queries and indexable data set points.
For all distances except R\'{e}nyi divergence we use 1K queries for each split, i.e., the total number of queries is 3K. Because R\'{e}nyi divergence is slow to compute,
we use only 200 queries per split (i.e., the overall number of queries is 600).

\FloatBarrier

\begin{figure}[!htb]
\centering
\includegraphics[width=0.75\textwidth]{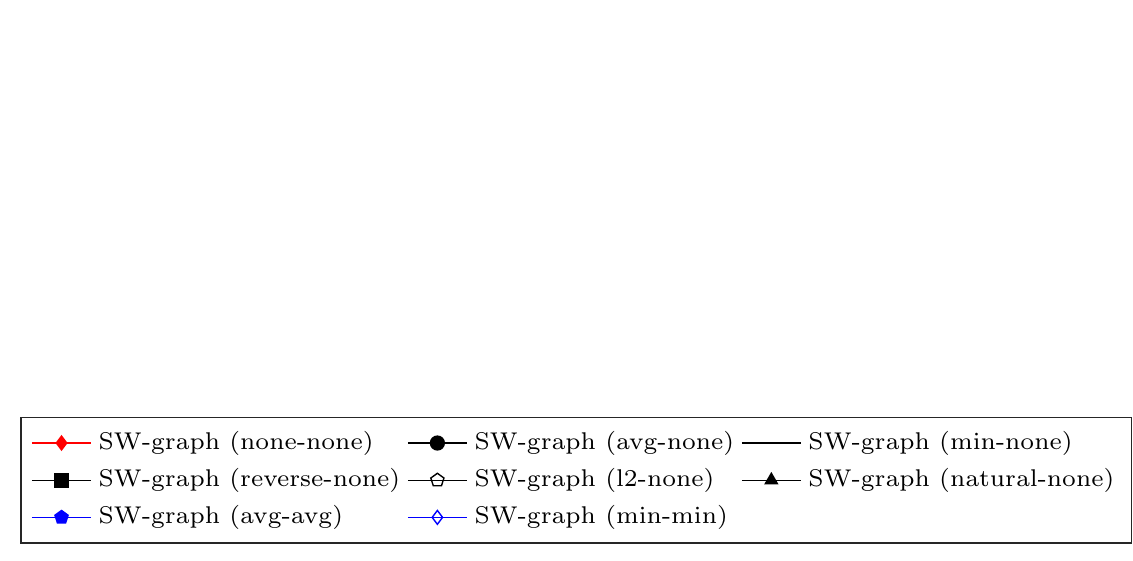}
\vspace{-1em}
\\
\subfloat[\scriptsize\label{PanelSymmFin_RCV8_ItakuraSaito}RCV-8 (Itakura-Saito)]{\includegraphics[width=0.310000\textwidth]{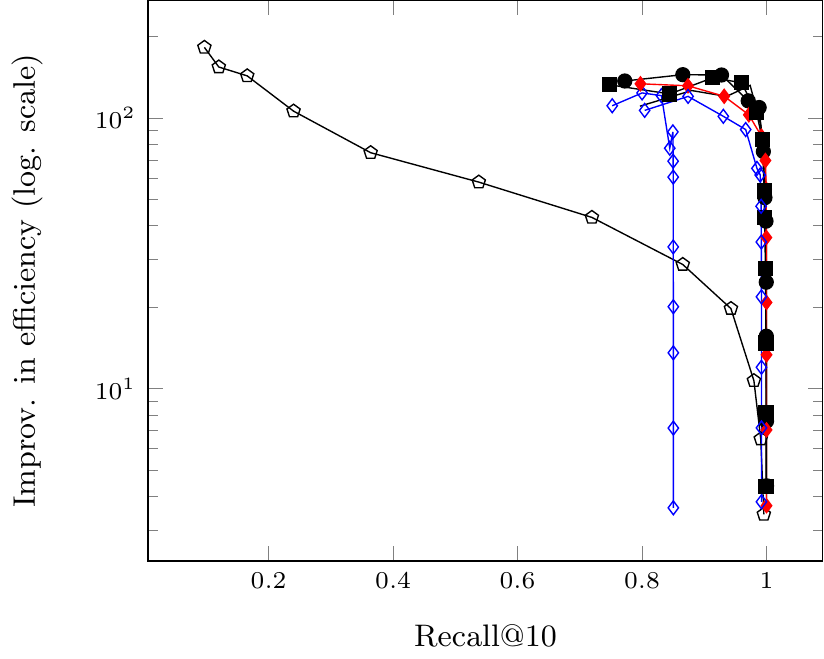}}
\subfloat[\scriptsize\label{PanelSymmFin_Wiki8_ItakuraSaito}Wiki-8 (Itakura-Saito)]{\includegraphics[width=0.310000\textwidth]{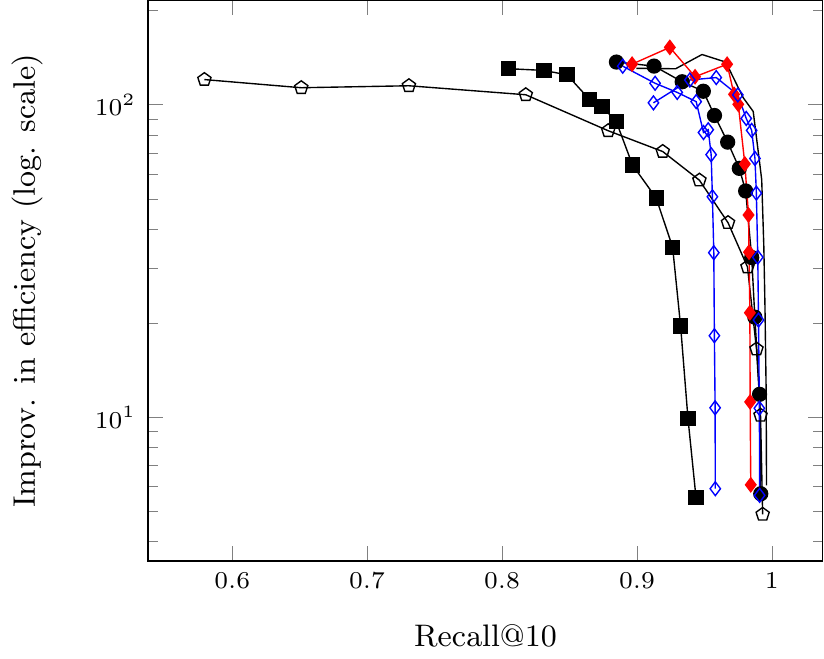}}
\subfloat[\scriptsize\label{PanelSymmFin_RandHist8_ItakuraSaito}RandHist-8 (Itakura-Saito)]{\includegraphics[width=0.310000\textwidth]{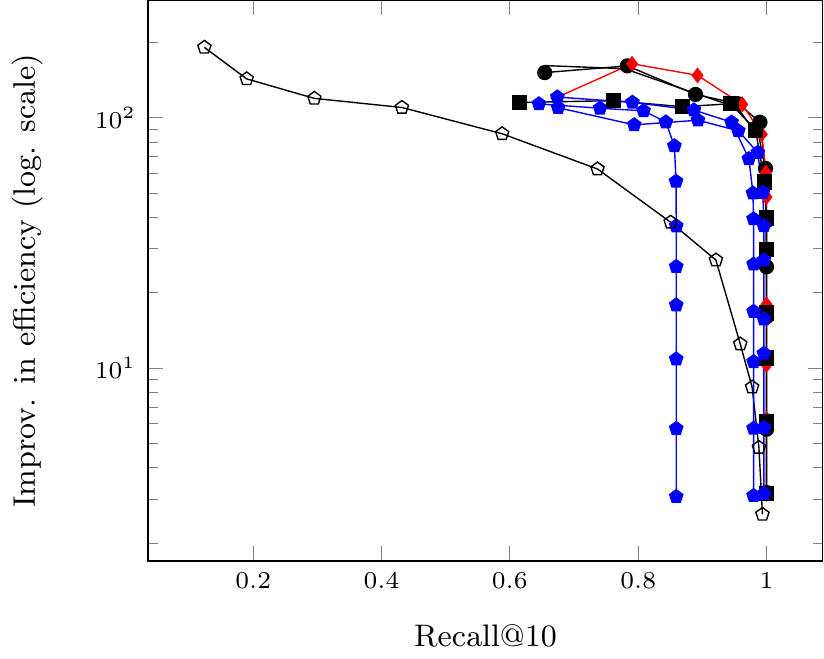}}
\\
\subfloat[\scriptsize\label{PanelSymmFin_RCV8_KLdiv}RCV-8 (KL-div.)]{\includegraphics[width=0.31\textwidth]{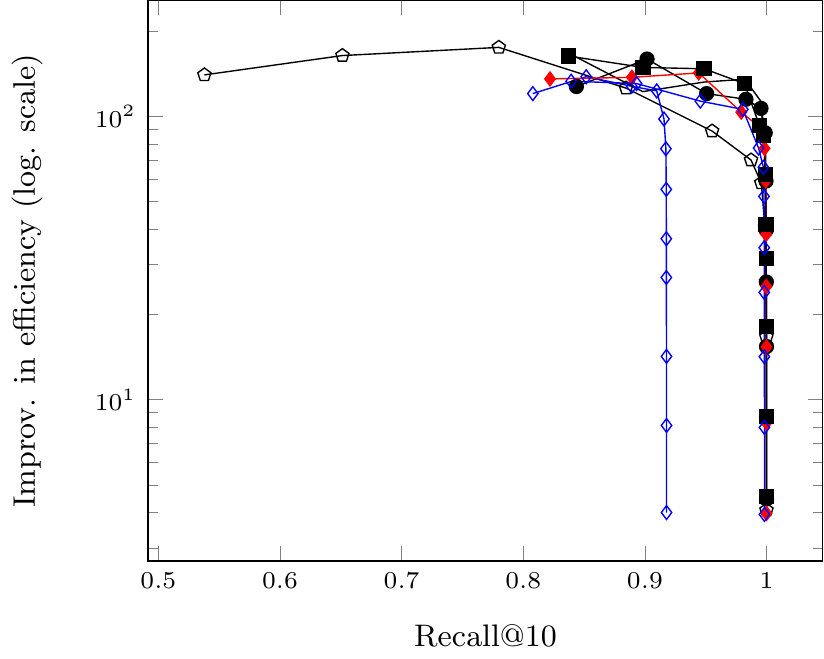}}
\subfloat[\scriptsize\label{PanelSymmFin_Wiki8_KLdiv}Wiki-8 (KL-div.)]{\includegraphics[width=0.31\textwidth]{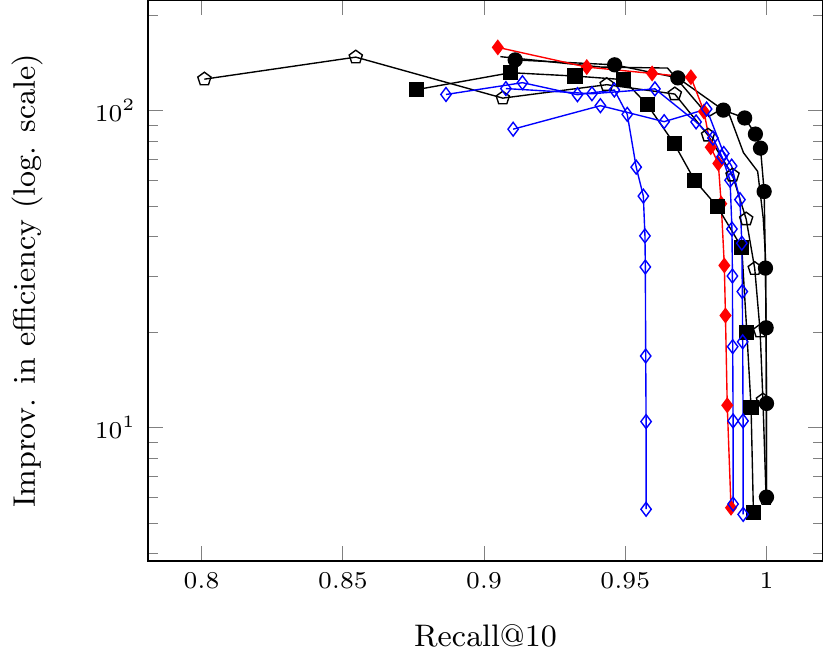}}
\subfloat[\scriptsize\label{PanelSymmFin_RandHist8_KLdiv}RandHist-8 (KL-div.)]{\includegraphics[width=0.31\textwidth]{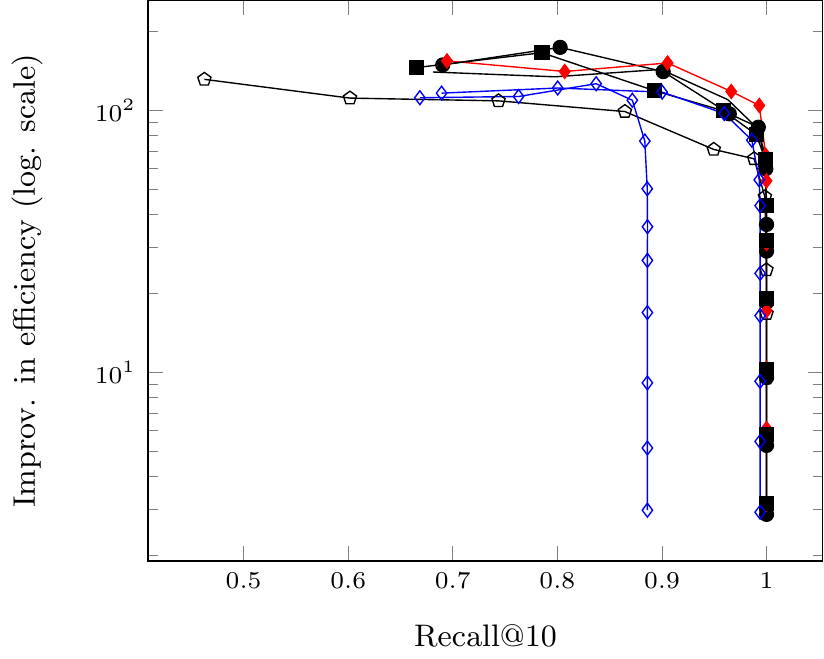}}
\\
\subfloat[\scriptsize\label{PanelSymmFin_RCV8_RenyiDiv025}RCV-8 (R\'{e}nyi div. $\alpha=0.25$)]{\includegraphics[width=0.310000\textwidth]{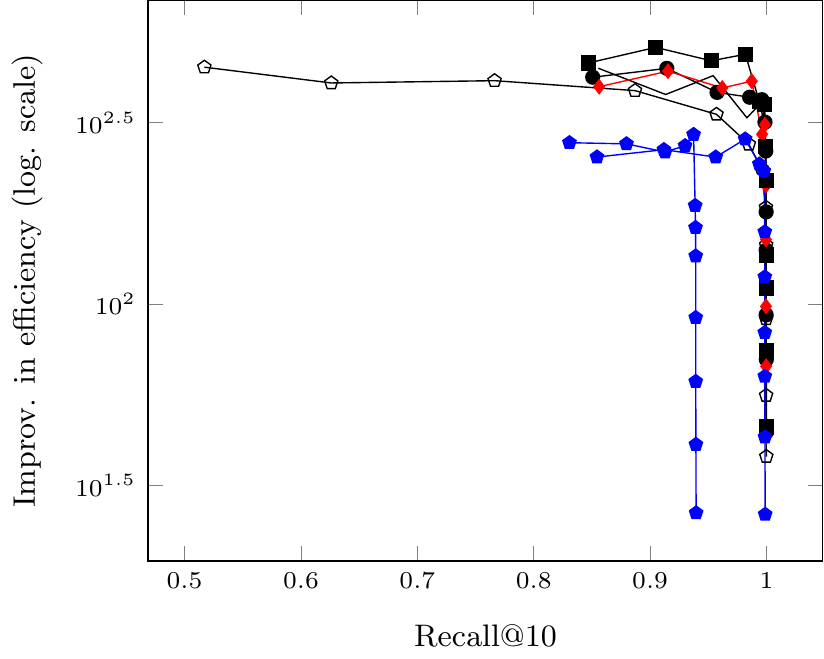}}
\subfloat[\scriptsize\label{PanelSymmFin_Wiki8_RenyiDiv025}Wiki-8 (R\'{e}nyi div. $\alpha=0.25$)]{\includegraphics[width=0.310000\textwidth]{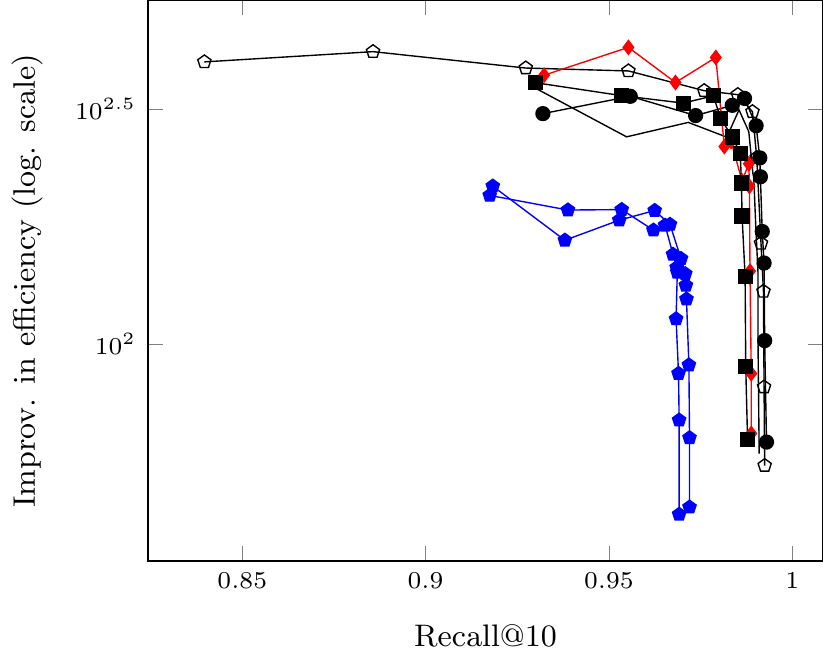}}
\subfloat[\scriptsize\label{PanelSymmFin_RandHist8_RenyiDiv025}RandHist-8 (R\'{e}nyi div. $\alpha=0.25$)]{\includegraphics[width=0.310000\textwidth]{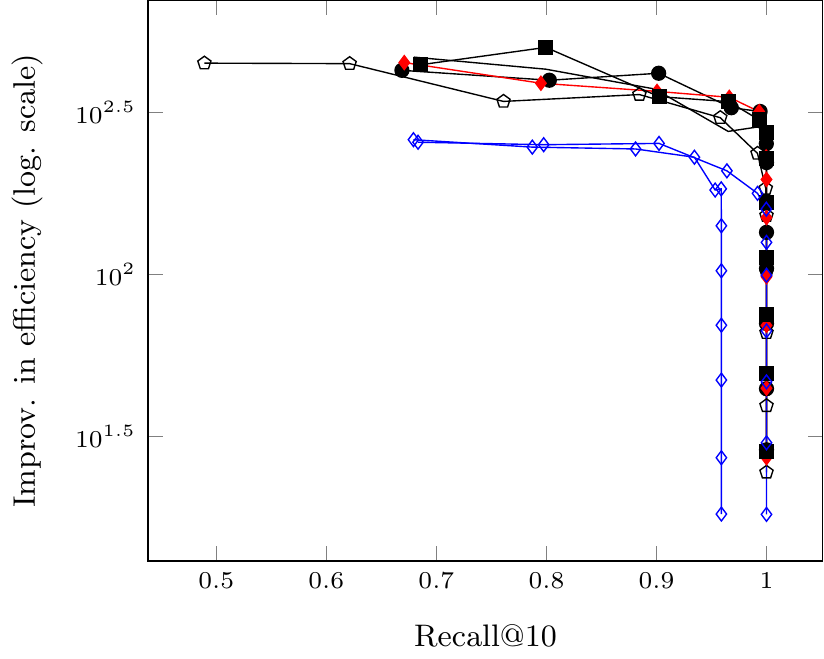}}
\\
\subfloat[\scriptsize\label{PanelSymmFin_RCV8_RenyiDiv075}RCV-8 (R\'{e}nyi div. $\alpha=0.75$)]{\includegraphics[width=0.310000\textwidth]{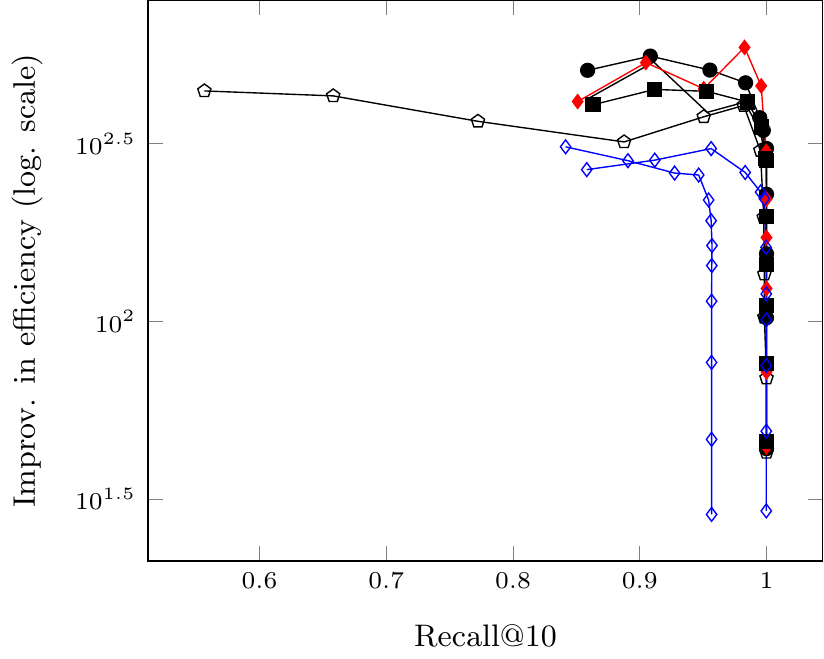}}
\subfloat[\scriptsize\label{PanelSymmFin_Wiki8_RenyiDiv075}Wiki-8 (R\'{e}nyi div. $\alpha=0.75$)]{\includegraphics[width=0.310000\textwidth]{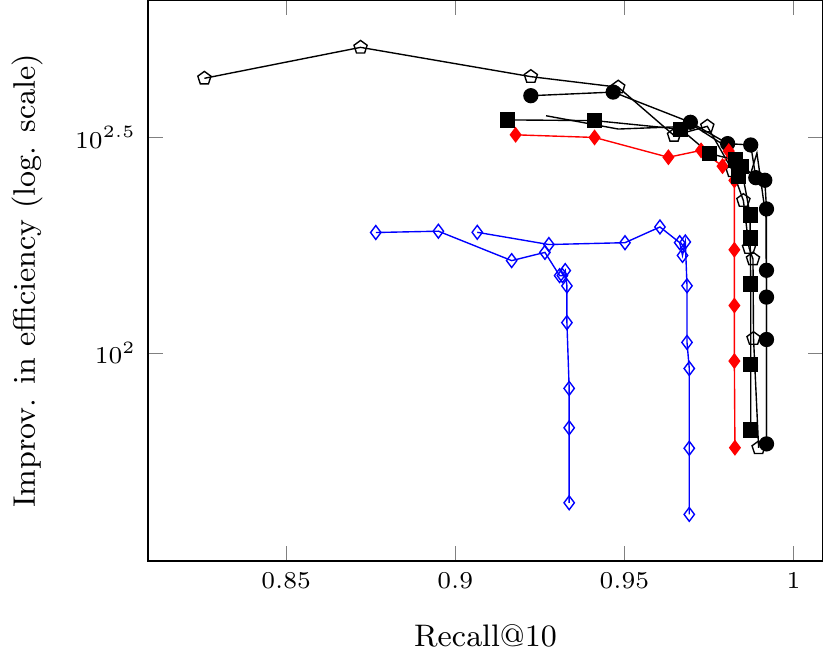}}
\subfloat[\scriptsize\label{PanelSymmFin_RandHist8_RenyiDiv075}RandHist-8 (R\'{e}nyi div. $\alpha=0.75$)]{\includegraphics[width=0.310000\textwidth]{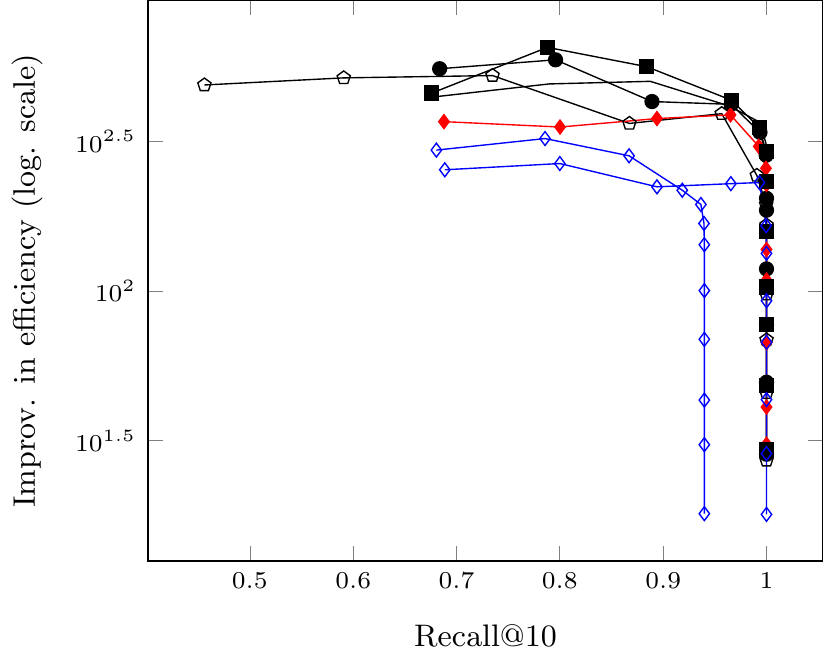}}
\\
\caption{\label{FigSymmFinal1} Efficiency/effectiveness trade-offs of symmetrization in 10-NN search (part I). The number of data points is at most 500K. Best viewed in color.}
\end{figure}

\begin{figure}[!htb]
\centering
\includegraphics[width=0.75\textwidth]{images/symm_final/legend_only_symm_final.pdf}
\vspace{-1em}
\\
\subfloat[\scriptsize\label{PanelSymmFin_RCV128_ItakuraSaito}RCV-128 (Itakura-Saito)]{\includegraphics[width=0.310000\textwidth]{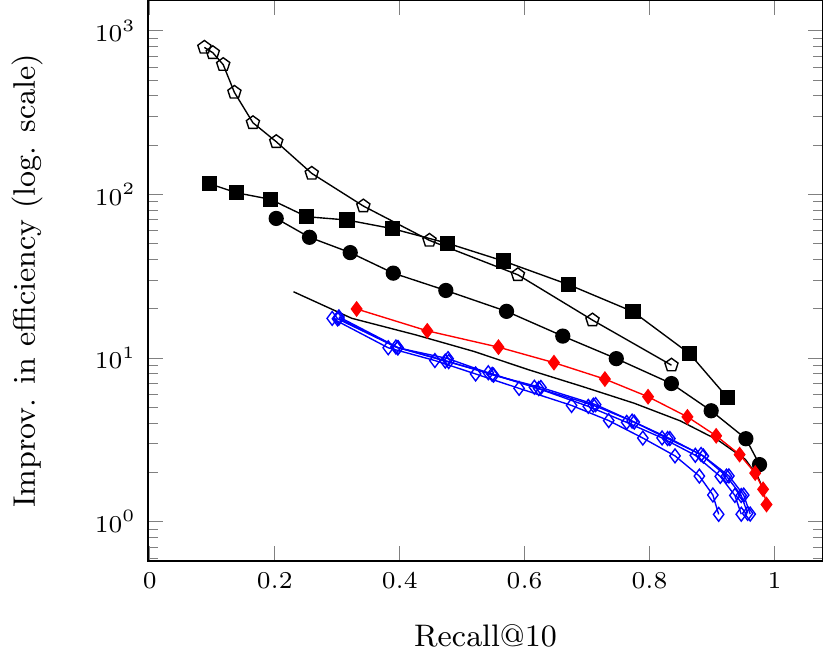}}
\subfloat[\scriptsize\label{PanelSymmFin_Wiki128_ItakuraSaito}Wiki-128 (Itakura-Saito)]{\includegraphics[width=0.310000\textwidth]{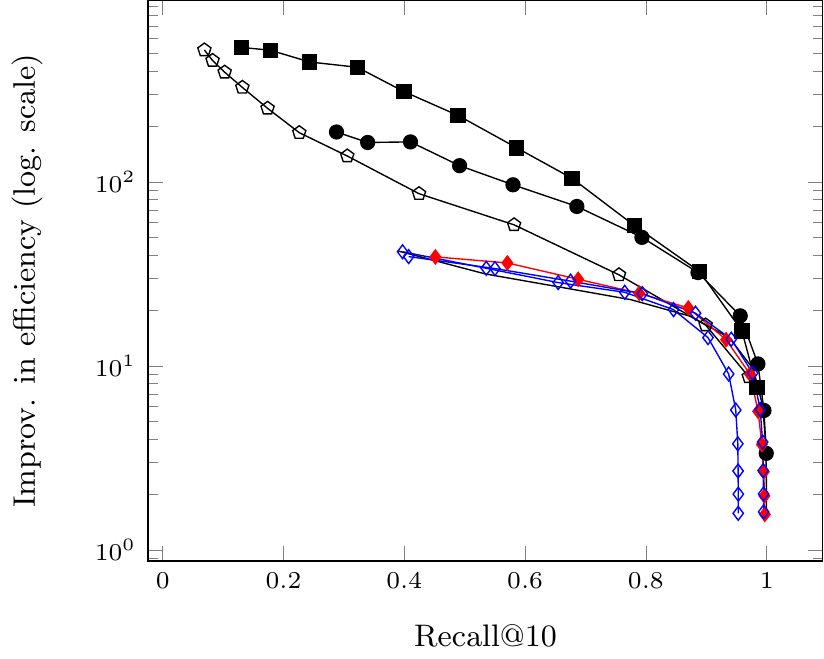}}
\subfloat[\scriptsize\label{PanelSymmFin_RandHist32_ItakuraSaito}RandHist-32 (Itakura-Saito)]{\includegraphics[width=0.310000\textwidth]{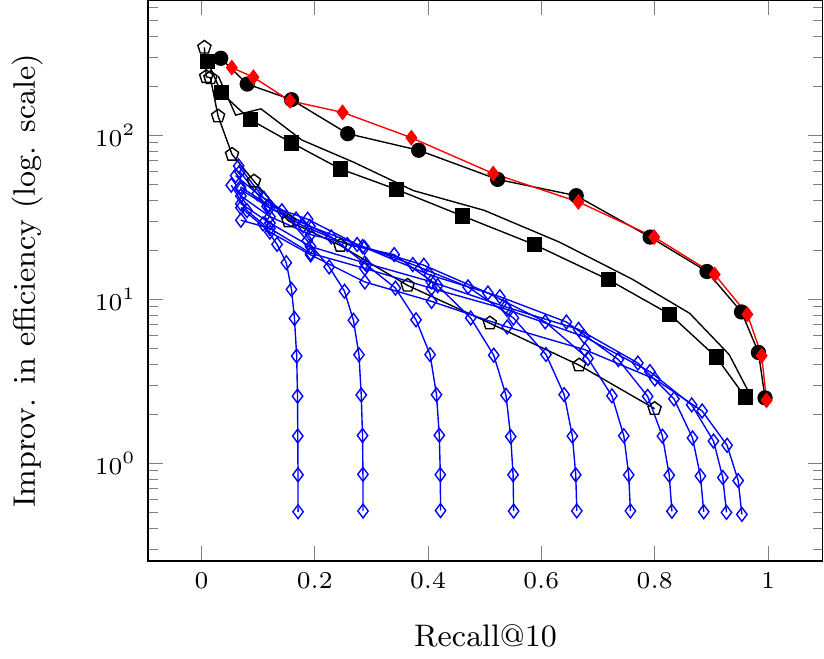}}
\\
\subfloat[\scriptsize\label{PanelSymmFin_RCV128_RenyiDiv025}RCV-128 (R\'{e}nyi div. $\alpha=0.25$)]{\includegraphics[width=0.310000\textwidth]{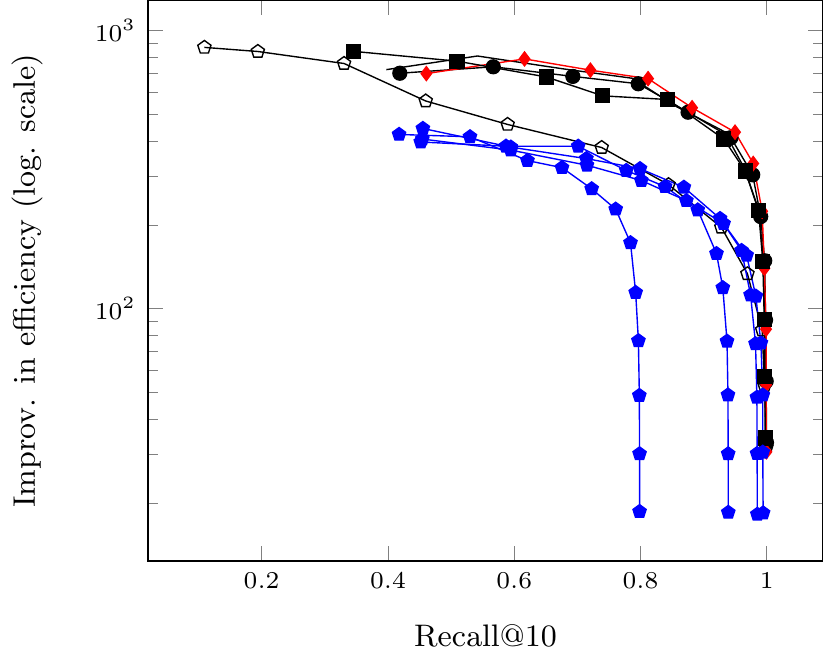}}
\subfloat[\scriptsize\label{PanelSymmFin_Wiki128_RenyiDiv025}Wiki-128 (R\'{e}nyi div. $\alpha=0.25$)]{\includegraphics[width=0.310000\textwidth]{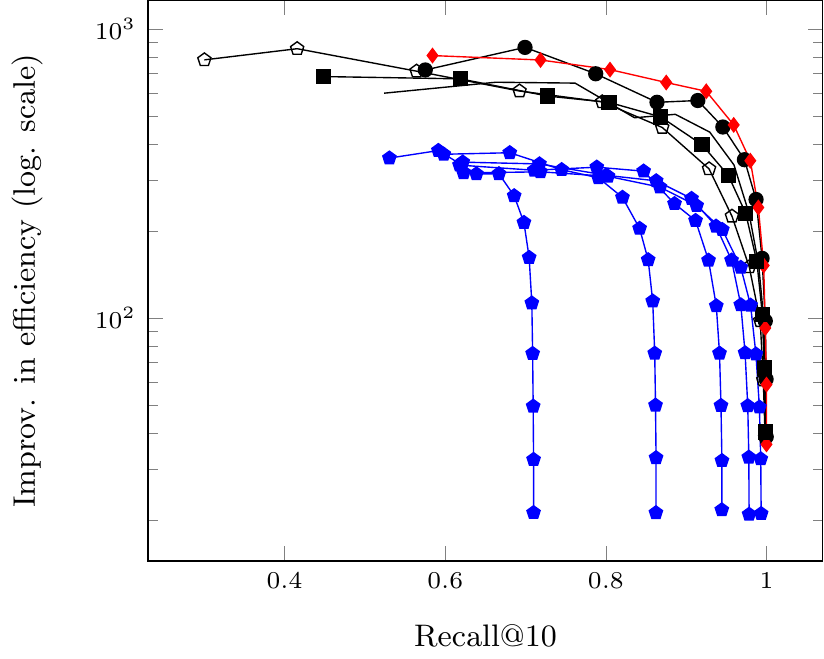}}
\subfloat[\scriptsize\label{PanelSymmFin_Manner_BM25}Manner (BM25)]{\includegraphics[width=0.310000\textwidth]{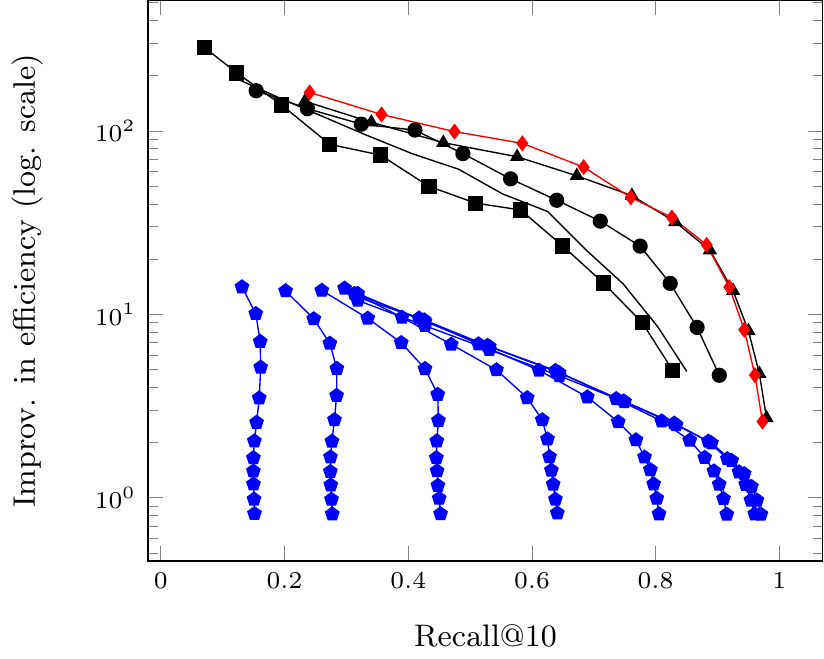}}
\\
\subfloat[\scriptsize\label{PanelSymmFin_RCV128_RenyiDiv075}RCV-128 (R\'{e}nyi div. $\alpha=0.75$)]{\includegraphics[width=0.310000\textwidth]{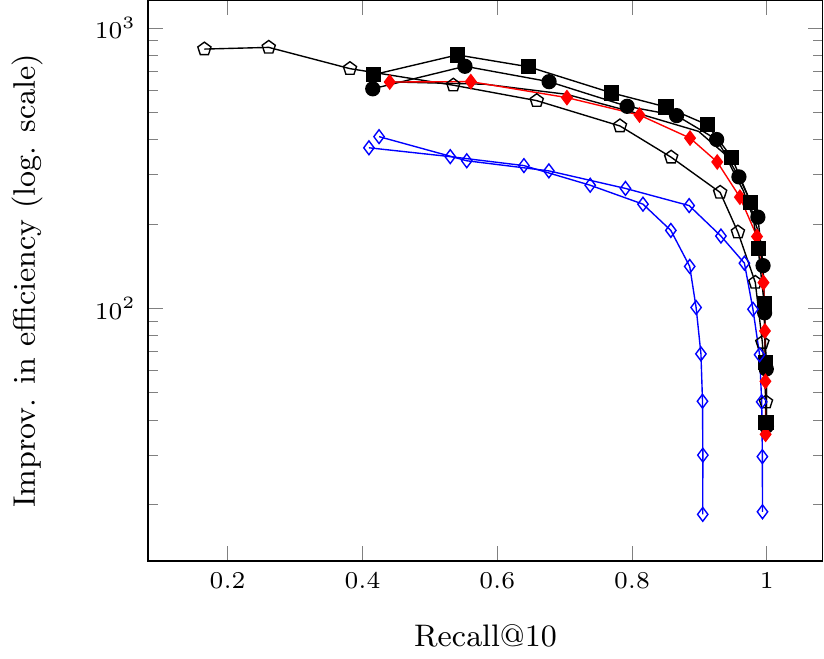}}
\subfloat[\scriptsize\label{PanelSymmFin_Wiki128_RenyiDiv075}Wiki-128 (R\'{e}nyi div. $\alpha=0.75$)]{\includegraphics[width=0.310000\textwidth]{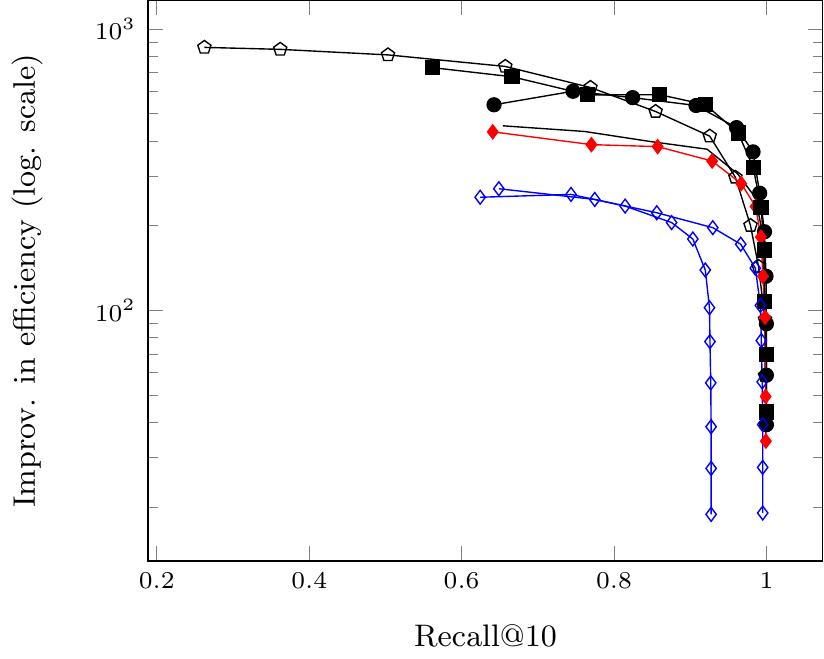}}
\subfloat[\scriptsize\label{PanelSymmFin_RandHist32_RenyiDiv075}RandHist-32 (R\'{e}nyi div. $\alpha=0.75$)]{\includegraphics[width=0.310000\textwidth]{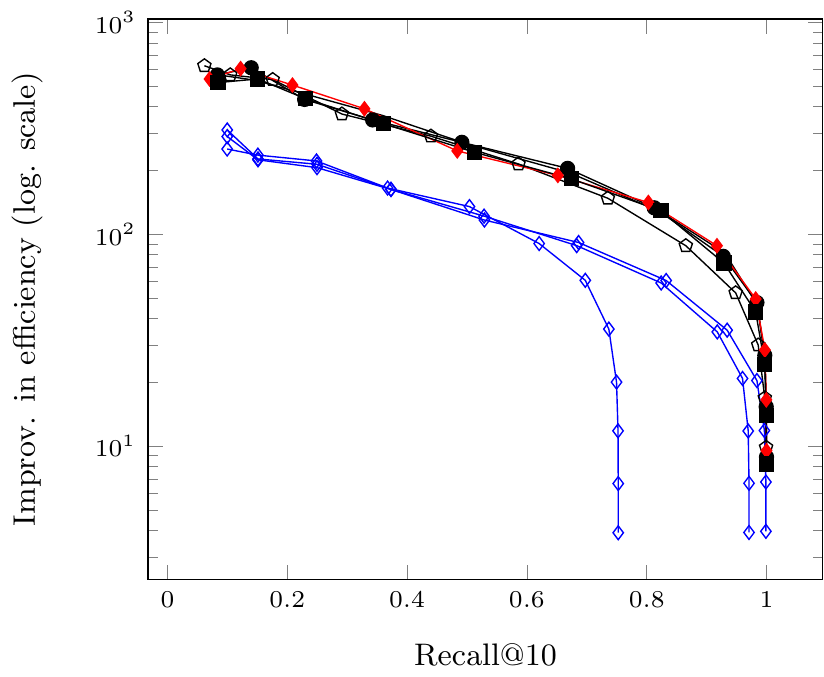}}
\\
\subfloat[\scriptsize\label{PanelSymmFin_RCV128_RenyiDiv2}RCV-128 (R\'{e}nyi div. $\alpha=2$)]{\includegraphics[width=0.310000\textwidth]{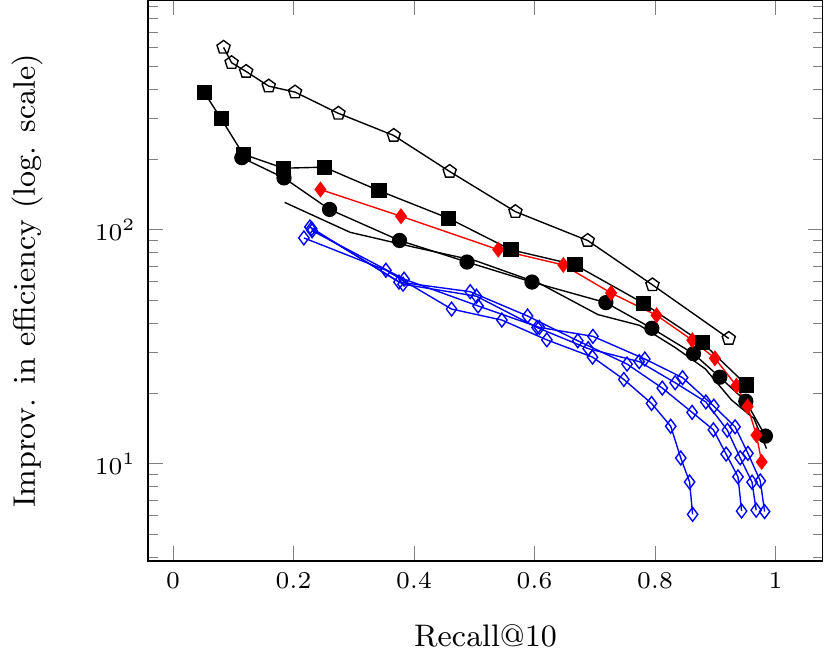}}
\subfloat[\scriptsize\label{PanelSymmFin_Wiki128_RenyiDiv2}Wiki-128 (R\'{e}nyi div. $\alpha=2$)]{\includegraphics[width=0.310000\textwidth]{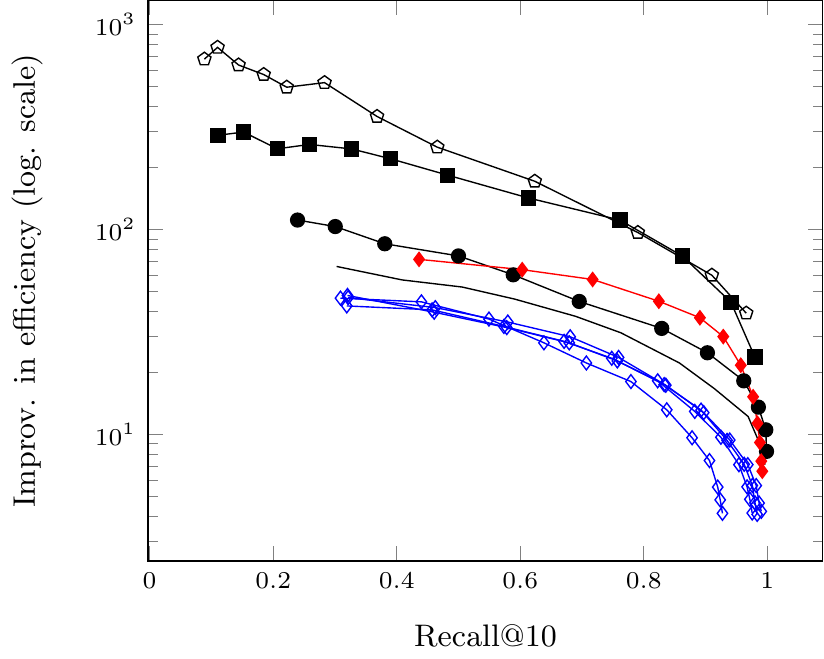}}
\subfloat[\scriptsize\label{PanelSymmFin_RandHist32_RenyiDiv2}RandHist-32 (R\'{e}nyi div. $\alpha=2$)]{\includegraphics[width=0.310000\textwidth]{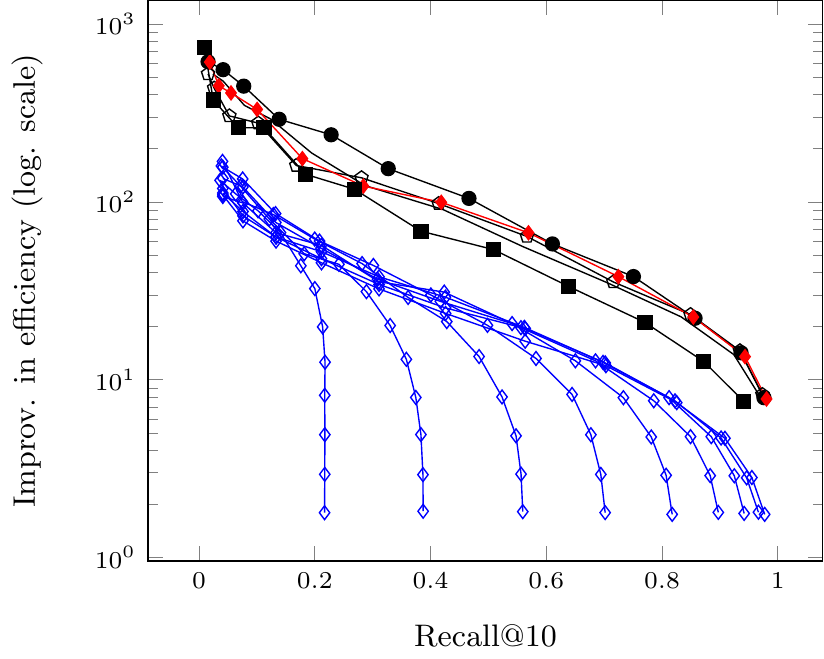}}

\caption{\label{FigSymmFinal2} Efficiency/effectiveness trade-offs of symmetrization in 10-NN search (part II). The number of data points is at most 500K. Best viewed in color.}
\vspace{-0.5em}
\end{figure}

\FloatBarrier

Experiments are carried out using a  \ttt{nmslib4a\_bigger\_reruns} branch\footnote{\url{https://github.com/nmslib/nmslib/tree/nmslib4a_bigger_reruns}} of NMSLIB \cite{SISAP2013}.
We did not modify the standard NMSLIB code for SW-graph: Instead, we created a new implementation
(file \ttt{small\_world\_rand\_symm.cc}).

In the second scenario,
we experiment with index- and query-time symmetrization in an actual indexing algorithm SW-graph rather than relying on the brute-force search.
This approach generates a \emph{final} list of $k$ nearest neighbors rather than $k_c$ candidates. \emph{No} further re-ranking is necessary.
We use two actual symmetrization strategies (the minimum- and the average-based
symmetrization) as well as two types of \emph{quasi}-symmetrization.
For the first quasi-symmetrization type, we build the proximity graph using the Euclidean distance between vectors.
The second quasi-symmetrization consists in building the proximity graph using the argument-reversed distance (see p.~\pageref{ParaSymmDesc}). 

We verified that \emph{none} of these quasi-symmetrization approaches would produce a better list of candidates
in the filter-and-refine scenario (where the brute-force search is used to produce a candidate list). 
For example, for Wiki-128 and KL-divergence, it takes $k_c=40$ candidates to exceed a 99\%
recall in a 10-NN search for the minimum-based symmetrization.
For the $L_2$-based symmetrization, it takes as many as $k_c=320$ candidates.
The results are even worse for the filtering based on the argument-reversed distance:
By using as many as $k_c=1280$ candidates we obtain a recall of only $95.6\%$.
It clearly does not make sense to evaluate these quasi-symmetrization methods
in the complete filter-and-refine scenario.
Yet, we need to check if it is beneficial to build the graph using
a distance \emph{different} from the original one.

\vspace{-0.5em}
\paragraph{Discussion of Results}
Experiments were run on a laptop (i7-4700MQ @ 2.40GHz with 16GB of memory).
Results are presented in Fig. \ref{FigSymmFinal1} (low-dimensional data) and Fig. \ref{FigSymmFinal2} (high-dimensional data).
These are efficiency-effectiveness plots: Recall@10 is shown on the x-axis,
improvement in efficiency---i.e., the speed up over the brute-force search---
is shown on the y-axis. Higher and to the right is better.
We test several modifications of SW-graph each of which has an additional marker in the form: \ttt{a-b},
where \ttt{a} denotes a type of index-time symmetrization and \ttt{b}
denotes a type of query-time symmetrization.
Red plots represent the original SW-graph, which is labeled as SW-graph (none-none).

Black plots represent modifications, where symmetrization is used only during indexing:
SW-graph (avg-none), SW-graph (min-none), SW-graph (l2-none), SW-graph (reverse-none), and SW-graph (natural-none).
The first two types of symmetrization are average- and minimum-based.
SW-graph (l2-none) is a quasi-symmetrization approach that builds the graph using $L_2$,
but searches using the original distance.
SW-graph (reverse-none) builds the graph using the reversed-argument distance,
but searches using the original distance.
SW-graph (natural-none) is a natural symmetrization of BM25 described by Eq.~(\ref{EqSymmNatural}),
which is used only for \emph{Manner}. 

Blue plots represent the case of full (both query- and index-time) symmetrization.
The index is used to carry out a $k_c$-NN search, which produces a list of $k_c$ candidates for further verification.
Depending on which symmetrization approach was more effective in the 
the first series experiments (with brute-force search), 
we use either SW-graph (min-min) or SW-graph (avg-avg),
which stand for full minimum- or average-based symmetrization.
Because we do not know an optimum number of candidate records,
we experiment with $k_c = k \cdot 2^i$ for successive integer values $i$.
The larger is $i$, the more accurate is the filtering step and the less efficient is retrieval.
However, it does not make sense to increase $i$ beyond the point where the filtering accuracy reaches 99\%.
For this reason, the minimum value of $k_c$ is $k$ and the largest value of $k_c$ is taken from Table~\ref{TableSymmMetrLearn}.
\vspace{-0.05em} 

For the remaining parameters of SW-graph we choose values that are known to perform well in other experiments: \ttt{NN}=15, \ttt{efConstruction}=100, and \ttt{efSearch} = $2^j$ for $0 \le j \le 12$.
Analogous to the first scenario (with brute-force search), we use 31 combination of data sets and distances.
In each test, we randomly split data (into queries and indexable data)
three times and average results over three splits.
\vspace{-0.05em} 

From Figures~\ref{FigSymmFinal1}-\ref{FigSymmFinal2},
we can see that in some cases there is little difference among best runs with the fully symmetrized
distance (a method SW-graph (min-min) or SW-graph (avg-avg)) the runs produced by methods
with true index-time symmetrization (SW-graph (min-none), SW-graph (avg-none)),
and the original unmodified search algorithm (SW-graph (none-none)).
Furthermore, we can see that there is often no difference between SW-graph (min-none), SW-graph (avg-none),
and SW-graph (none-none).
However, sometimes all fully-symmetrized runs (for all values of $k_c$) are noticeably less efficient (see, e.g., Panels \ref{PanelSymmFin_Wiki8_RenyiDiv025} and \ref{PanelSymmFin_Wiki8_RenyiDiv075}).
This difference is more pronounced in the case of high-dimensional data.
Here, full symmetrization leads to a substantial (up to an order of magnitude)
loss in performance in most cases.

Effectiveness of index-time symmetrization varies from case to case and there is no definitive winner.
First, we note that in four cases index-time symmetrization is beneficial (Panels
\ref{PanelSymmFin_RCV128_ItakuraSaito}, \ref{PanelSymmFin_Wiki128_ItakuraSaito},
\ref{PanelSymmFin_RCV128_RenyiDiv2}, \ref{PanelSymmFin_Wiki128_RenyiDiv2}).
In particular, in Panels
\ref{PanelSymmFin_RCV128_ItakuraSaito}, \ref{PanelSymmFin_Wiki128_ItakuraSaito}, \ref{PanelSymmFin_Wiki128_RenyiDiv2}, there is an up to 10$\times$ speedup.
Note that it can sometimes be achieved by using an argument-reversed distance (Panels \ref{PanelSymmFin_RCV128_ItakuraSaito}, \ref{PanelSymmFin_Wiki128_ItakuraSaito}) or $L_2$ (\ref{PanelSymmFin_Wiki128_RenyiDiv2}).
This a \emph{surprising} finding given that these quasi-symmetrization approaches do not perform well in the re-ranking--filter-and-refine---experiments.
In particular, for $L_2$ and Wiki-128 reaching a 99\% recall requires $k_c=640$ compared to $k_c=80$
for min-based symmetrization.
For the Itakura-Saito distance and data sets RCV-128 and Wiki-128,
it takes $k_c \le 80$ to get a 99\% recall.
However, using the argument-reversed distance, we do not even reach the recall of 60\% despite using a large $k_c=1280$.
It is worth noting, however, that
in several cases using argument-reversed distance at index time leads to substantial degradation in performance
(see, e.g., Panels \ref{PanelSymmFin_Wiki8_ItakuraSaito} and \ref{PanelSymmFin_Manner_BM25}).

To conclude the section, we emphasize
that in all cases the best performance is achieved using either the unmodified SW-graph or the SW-graph
with an index-time proxy distance.
However, there is not a single case where performance is improved by using the fully symmetrized distance (at both indexing and querying steps).
Furthermore,
in three especially challenging cases:
Itakura-Saito distance with RandHist-32,  R\'{e}nyi divergence with RandHist-32,
and BM25 with Manner,
SW-graph has excellent performance.
In all three cases (see Panels~\ref{PanelSymmFin_RandHist32_ItakuraSaito},
\ref{PanelSymmFin_RandHist32_RenyiDiv2},\ref{PanelSymmFin_Manner_BM25}),
there is more than a 10$\times$ speed up at 90\% recall compared to 
the brute-force search.
Note that in these three cases data is substantially non-symmetric:
Depending on the case,
to accurately retrieve 10 nearest neighbors with respect to the original metric,
it requires to obtain 1-5K nearest neighbors using its symmetrized variant
(see Table~\ref{TableSymmMetrLearn}). 
Thus, in these challenging cases, a brute-force filter-and-refine symmetrization solution would be particularly ineffective or inefficient whereas SW-graph has strong performance.
\newpage 

\section{Conclusion}
We  systematically evaluate effects
of distance metrization, symmetrization and quasi-symmetrization on 
performance of brute-force and index-based \knn\ search (with a graph-based
retrieval method SW-graph).
Unlike  previous work \cite{Naidan2015,ponomarenko2014comparative}
we experiment with substantially  non-symmetric distances.
Coercion of the non-metric distance to a metric space
leads to a substantial performance degradation.
Distance symmetrization causes a lesser performance loss.
However, in all the cases a full filter-and-refine symmetrization 
is always inferior to either applying the graph-based
retrieval method directly to a non-symmetric distance
or to building an index (which is a neighborhood graph)
with a modified, e.g. symmetrized, distance.
Quite surprisingly, sometimes
the best performing index-time distance is neither
the original distance
nor its symmetrization.
This observation motivates a new line of research
of designing index-specific graph-construction distance functions.
\vspace{-0.5em}

\paragraph{Acknowledgments}
This work was done while Leonid Boytsov was a PhD student at CMU.
Authors gratefully acknowledge the support by the NSF grant \#1618159.

%
%
%

\end{document}